\documentclass[twoside]{ilcws10}
\usepackage[latin1]{inputenc}
\usepackage[dvips]{graphicx,epsfig,color}
\usepackage{wrapfig,rotating}
\usepackage{amssymb,amsmath,array}
\graphicspath{{./figure/}}
\newcommand{\etal}{{\it et~al.}}
\begin{document}
\title{
  Photon detection efficiency of  
Geiger-mode avalanche photodiodes 
} 
\author{S.~Gentile$^1$, E.~Kuznetsova$^2$, F.~Meddi$^1$
\vspace{.3cm}\\
1- Universit\`{a} degli Studi di Roma "La Sapienza",\\
Piazzale Aldo Moro 5, 00185 Roma, Italy.
\vspace{.1cm}\\
2- DESY,\\
Notkestra$\ss$e 85, 22607, Hamburg, Germany. \\
}

\maketitle

\begin{abstract}

 The photon detection efficiencies 
of multi-pixel Geiger-mode avalanche photodiodes manufactured by different producers are estimated.
A new  fit method of the response spectra to low-intensity light, taking into account after-pulse and  cross-talk effects is proposed to  yield
 the initial number of photons.
The value of photon detection efficiency is calculated
using a calibrated photodetector as a reference.
\end{abstract}

\section{Introduction}\label{sec:intro}
Multi-pixel Geiger-mode avalanche photodiodes (G-APDs) are  solid-state
photodetectors based on a 
technology developed since
early  1990's 
\cite{Bondarenko:1998cj, Buzhan:2003ur,
GolovinSaveliev:2004nima, Dolgoshein:2006nima,
HAMAMATSU:IEEE2006, IRST:2007nima,Bonanno:2009ab,Danilov:2009nima,Danilov:2007nima}.  
Their gain is typically  10$^6$ per photoelectron.
 Their main features are: the small size, the  efficiency comparable to the one of vacuum photomultipliers,
the insensitivity to magnetic field, a low bias voltage and a reasonable price.
Currently, the G-APDs  find an application  in many fields of physics,
High-Energy, 
 Neutrino, 
  Astroparticle and Medicine.

 This paper  describes the results of a comparative study performed  in the  same experimental  conditions, on  different G-APDs,
 to  allow  an optimal choice  with a view to a given application. We refer to elsewhere for a  complete description of measurements and analysis methods ~\cite{Gentile:2010nuovo}.

\section{General principle of the G-APD operation and its characteristics}\label{sec:principle}
A G-APD consists of a large number of identical microcells (silicon diodes, pixels) with a common anode. The microcells are located on a common substrate with a typical size of $\sim 1\times 1~\rm{mm}^2$.
Once photoelectron are produced  an electron avalanche multiplication may occur under  a reverse bias, $\rm{V_{bias}}$, above the breakdown voltage, $\rm{V_{brd}}$ and the charge induced during
the avalanche discharge is proportional to the overvoltage ($\rm{V_{over}} =\rm{V_{bias}}$ -$\rm{V_{brd}}$).
 The number of fired pixels  is  proportional to the initial number of photoelectrons, as long as 
this is on average less than one per pixel.
This implies a  low-intensity light source in the measurement setup.
The gain (g) is determined by the charge released in one pixel.
\par Other  characteristics of  the  G-APD response to  light deserve a few words. 
\par  After-Pulses (AP) appear  when the quenching (mechanism to stop the avalanche process by temporally reducing the cell voltage) doesn't 
drain all the charge in the sensitive area, and  the cell  fires again  shortly after  the original pulse.
\par Cross-talk takes place when  in a Geiger discharge, some of the electrons generated in the avalanche process 
 reach the adjacent pixel, where another Geiger discharge might be triggered. This results in a pulse with a larger amplitude (nearly a factor two)
and in a distortion of the linear response of the device.

Every detector can only convert a certain percentage of incident photons in signals. This G-APD  efficiency
 is given as product of three factors:
\begin{equation}
 \rm{PDE = QE}\times \epsilon_{\rm geom} \times\epsilon_{\rm Geiger}, 
 \label{eq:PDE}
\end{equation}
\noindent where QE is the G-APD quantum efficiency, function of the incident  photon wavelength, $\lambda_{\rm w} $;
$\epsilon_{\rm geom}$ is a geometrical factor (fill factor) indicating  which fraction of the device is sensitive to photons; 
$\epsilon_{\rm Geiger}$ is the probability to trigger a Geiger discharge and is function of $\rm{V_{bias}}$. PDE decreases with the pixel size. 

\par  Then, the G-APD response can be written as a product of few factors:
 
  \begin{equation}
\rm  A =   N_{\gamma}^{in} \times {\rm PDE}\times g  \times (1+\varepsilon)\times(1+P_{AP}),  
 \label{eq:PDE1}
\end{equation} 
\noindent where $\rm N_{\gamma}^{in}$ is   the number of incident photons; g,  the gain; $\varepsilon$, the cross-talk probability and
 $\rm P_{AP}$, the after-pulse probability. 
It should be noticed that in the above equation, g, PDE, $\varepsilon$ and $\rm P_{AP}$ are  all increasing
 with  $\rm{V_{bias}}$, implying a complex dependence of  the G-APD response on the bias voltage.

\section{Measurement setup}\label{sec:setup}

\begin{wrapfigure}{r}{0.5\columnwidth}
\centerline{\includegraphics[width=0.45\columnwidth]{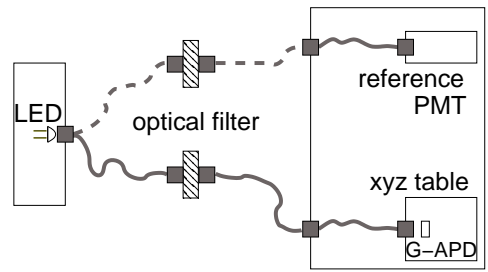}}
\caption{General scheme of the PDE measurements. }
\label{fig:setup1}
\end{wrapfigure}

Figure \ref{fig:setup1} shows the  scheme of the PDE measurements.
The light from a Light-Emitting Diode (LED) operating in a pulse-mode with a Waveform Generator (AGILENT 33220A) was
delivered to an optical filter. The latter
corresponded to the peak  wavelength of the LED and  had a bandwidth of  $\pm $ 3 nm. 
By changing the LED, a few wavelengths in  the 380 $\div$ 650 nm range  were investigated.
For low intensity sources (380 nm, 400 nm, 565 nm) the LED light was not filtered.

The filtered light was routed to a light-tight thermostabilized box  with inside two photo-detectors: the G-APD and
the reference photodetector.
The box temperature  was controlled at $\pm \ 0.1\ ^{\circ}\rm{C}$ during the data taking.
As a reference photodetector a  PhotoMultiplier Tube (PMT) H5783P  produced  and calibrated (for this reference) by HAMAMATSU  was used.
 For each  wavelength $\lambda_{\rm w}$ from the efficiency of the reference photodetector  for the peak wavelength, as provided by the manufacturer,
is derived  the efficiency of the reference device, $\epsilon_{\rm eval}^{\rm ref}$, for   our light source, taking in account the  filter bandwidth or the LED full width half maximum.

The signal from G-APD is read out with a charge-sensitive preamplifier 
and digitized with an integrating ADC (CAEN QADC V792N). The LED pulse of about $6\;\rm{ns}$ 
duration and the  ADC gate of about $65\;\rm{ns}$ width were synchronized by means of 
a common trigger.
The integration time was long enough to contain most of G-APD signal.
In case of the reference photodetector the direct signal from 
PMT was amplified with an external amplifier (NIM ORTEC 450) and digitized with the same ADC.
Due to a high sensitivity to  voltage variations, the  G-APD bias voltage was provided by a high precise and stable in time voltage source (Keithley 6487/E).

\section{Samples}\label{sec:samples}

The measurements refer to the samples listed in Tab. \ref{tab:samples}. The following
operating conditions were chosen:
\begin{itemize}
\item HAMAMATSU produced Multi-Pixel Photon Counter S10362-11-025U~\cite{hamamatsu} operated at
      ${\rm V_{bias}}$ = 71.32 V corresponding to ${\rm {V_{over}}} \simeq $ 2.3 V.

\item CPTA   produced SiliconPhotoMultiplier 
operated at  $\rm{V_{bias}}$ = 32.5 and 32.7 V corresponding to ${\rm {V_{over}}}\simeq 2.3$ V and  2.5 V.
The parameter dependence on voltage has been studied in the range  between 31.5 and 33.3 V.  

\item IRST    produced SiliconPhotoMultiplier operated at $\rm{V_{bias}}$ = 32 V corresponding to ${\rm {V_{over}}}\simeq 1.5 $ V.
\end{itemize}
The temperature at data taking was  T $ \sim 23.7 \ ^{\circ}\rm{C}$ stabilized at $\pm \ 0.1 \ ^{\circ}\rm{C}$ during each measurement
and with a spread $\pm \ 0.3 \  ^{\circ}\rm{C}$ during all period of data taking.
 
\begin{table}[h]
\centering
{
   }
  \begin{tabular}{|l|r|r|r|r|}
    \hline
 Sample      &  Type   & Photosensitive area & Number of pixels &Pixel pitch    \\ 
                  &         &  [$\rm mm^2$]         &                 & [$\rm \mu  m^2$]      \\
     \hline
      HAMAMATSU    &S10362-11-025U & 1x1            &        1600&        25x25 \\
      \hline
       CPTA[1]      &143            &1.028[2]&         556&        43x43 \\ 
\hline
       IRST[3]      &2007 prod      & 1x1           &          400&        50x50 \\
\hline
  \end{tabular}
\caption{Specification of measured samples. From left: Manufacturer, type, photosensitive area, number of pixels and their pitch.
           [1] Sample kindly provided by Prof. M. Danilov, 
           [2] Sensitive area  has octagonal shape,
           [3] Sample kindly provided by Dr. C. Piemonte (Fondazione Bruno Keller).}
  \label{tab:samples}
\end{table}

\section{Measurement strategy and fit procedure}\label{sec:Fit}

As described in Sec.~\ref{sec:setup} the  PDE measurement strategy is based
on the comparison of the effective number of photons detected from reference detector and  G-APD.

 To accomplish this task,  noise and  light (signal)   response spectra  of G-APD and PMT
to a low number of photons have been  measured and fitted, as  described in the following.

The number of photons, n, detected per  light pulse is expected to be Poisson distributed, with mean value $\lambda$. 
This distribution 
has to be convoluted with a gaussian distribution describing the experimental resolution. The charge x measured by the ADC
is expected to have a frequency distribution, N(x):

\begin{eqnarray}
\rm N(x) 
=N\times \left(\rm{Poisson(\lambda) \otimes \sum_{n\geq 0} \rm{Gauss}(\mu_n,\sigma_n})\right), \nonumber
\label{eq:eq1}
\end{eqnarray}
or more explicitly:
\begin{equation}
\rm N(x)=N\times e^{-\lambda} \sum_{n\geq 0}\left(\frac{\lambda^n}{n!}\times \frac{1}{\sigma_n}
exp(\frac{-(x-\mu_n)^2}{2\sigma_n^2}
 \right).
\label{eq:eq3}
\end{equation}

\noindent 

 In the ideal case n photons  correspond to a charge  $\rm \mu_n=\mu_0+n\times \rm{g}$,
where $\mu_0$ is the pedestal position and g is the gain factor.
 In presence of a finite resolution
the distribution of charge follows a gaussian distribution around each  $\rm \mu_n$ with a width $\sigma_n$,
$\rm \sigma_n^2=\sigma_0^2+n\times\sigma_1^2$, where $\sigma_0$ is
the electronic noise  (pedestal) width
 and $\sigma_1$ is  the single photon width. 
N is a normalization factor.

\begin{figure}[h]
\begin{center}
\includegraphics[width=0.45\linewidth,clip]{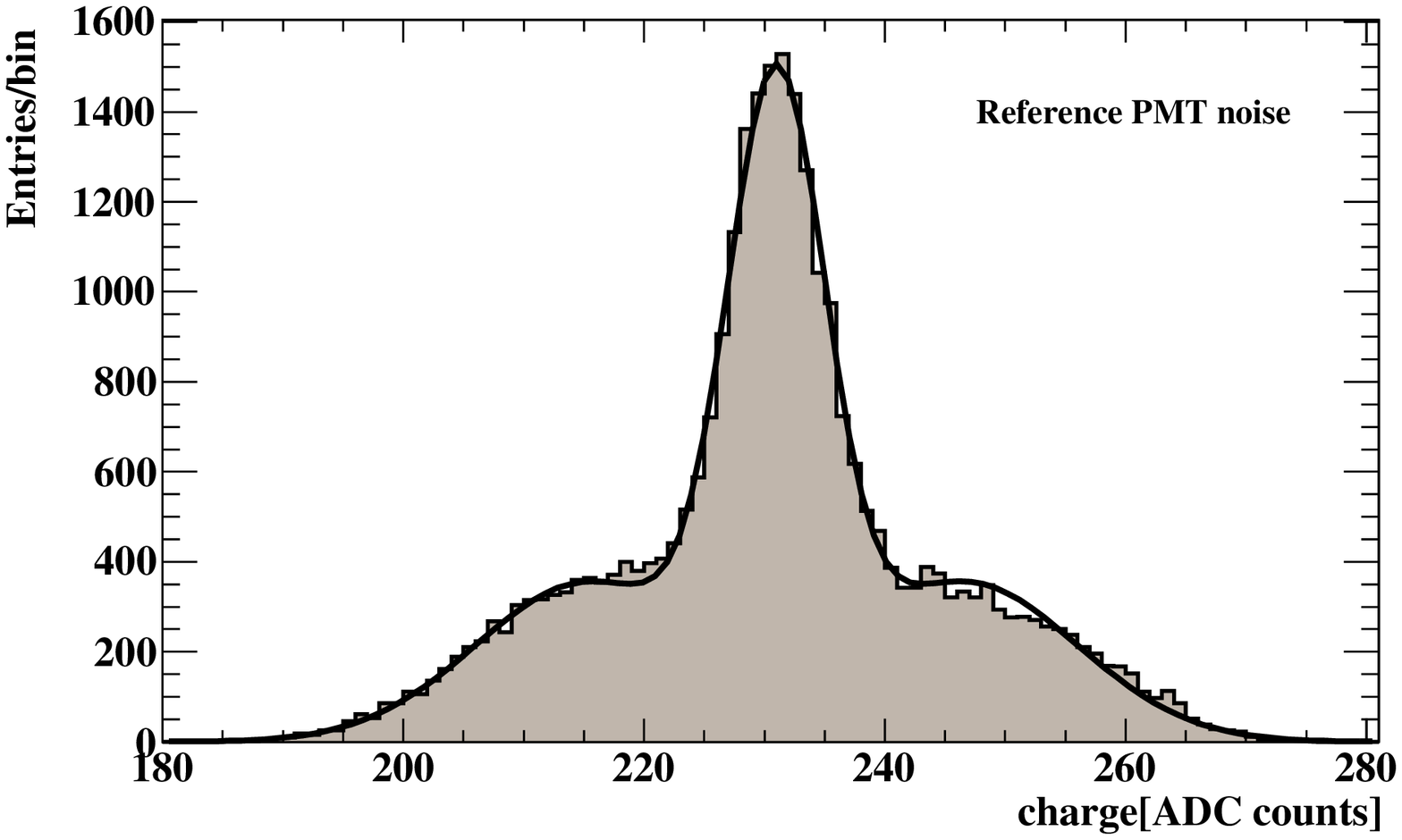}
\includegraphics[width=0.45\linewidth,clip]{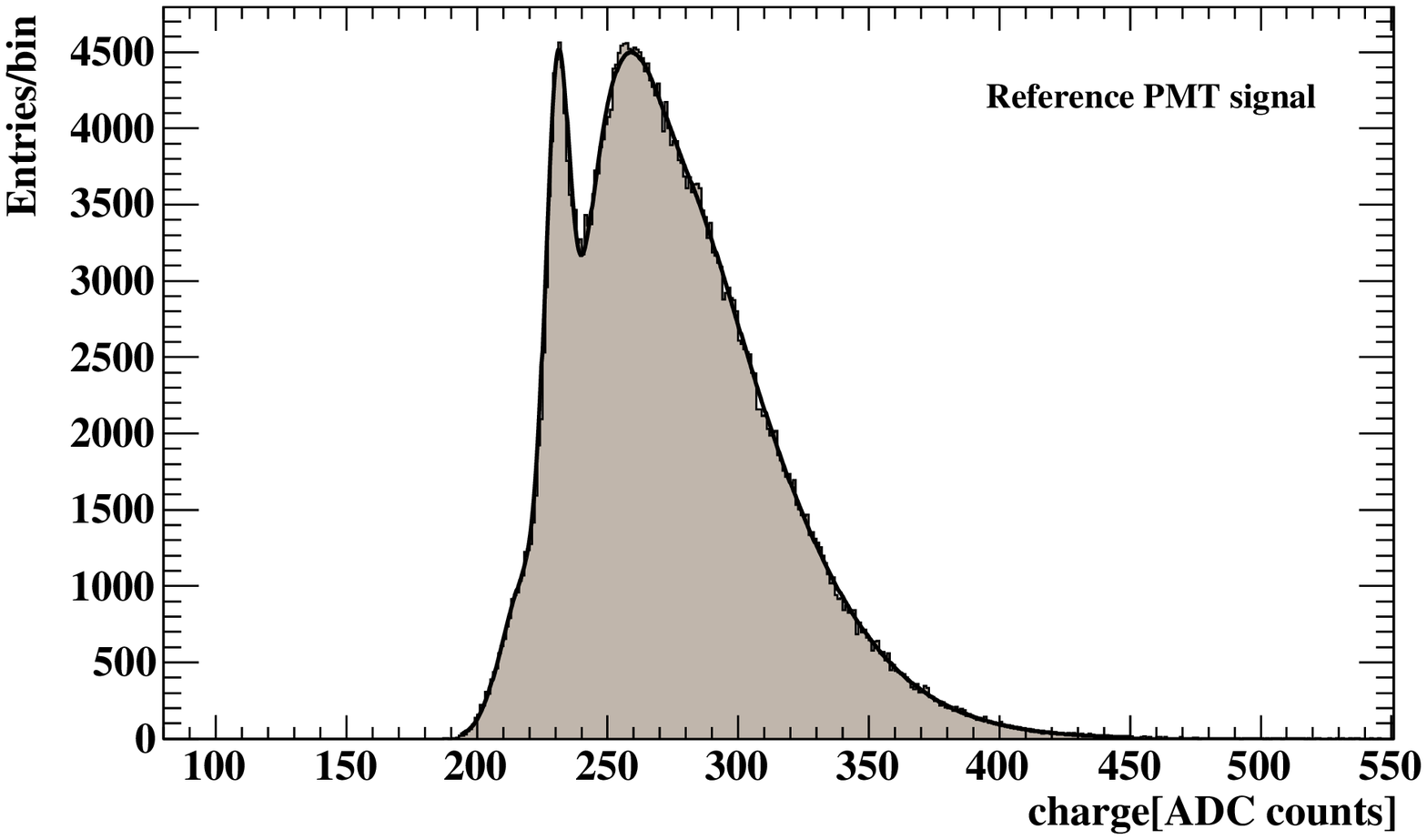}
\end{center}
\caption{
Reference PMT: noise  (left) and  signal (right) spectra.
The signal is  obtained  by short, low-intensity LED  flashes  at $\lambda_{\rm w}$ = 450 $\pm$ 3 nm.
The continuous lines are the fit results (see text).
From the fit the mean number of detected  photons  is obtained, $\rm N_{\gamma}^{PMT}=1.71 \pm 0.02 $ ($\epsilon_{\rm eval}^{\rm ref} $= 20.33 ~$^{+0.27}_{-0.28}$\%, see text).  Binwidth=1. }
\label{fig:refPMT}
\end{figure}
\paragraph{Reference PMT fit procedure}\label{sec:PMTFit}

 In the PMT case, it should be noticed  that due to additional electrical noise caused by the external amplifier, the
no-light spectrum (pedestal) is  described as
superposition of three gaussian peaks, two of which are considered to be
symmetric. Figure~\ref{fig:refPMT}~(left) shows the pedestal fitted with the
sum of three gaussians. Figure~\ref{fig:refPMT}~(right) shows the signal
 fitted with  Eq.~\ref{eq:eq3}  modified  to include the  noise shape.

Using the mean number of detected photons, $\lambda=\rm N_{\gamma}^{PMT}$, as obtained from the above fit, and the PMT
efficiency, 
the mean number of photons delivered by the optical system to 
the photodetector surface per one LED pulse was estimated. 
\par In the example of Fig.~\ref{fig:refPMT}, at $\lambda_{\rm w}= 450$ nm, 
 it is $\rm N_{\gamma}^{PMT}=1.71 \pm 0.02 $. 
 Taking into account the  efficiency for this $\lambda_{\rm w}$,  $\epsilon_{\rm eval}^{\rm ref} $ = 20.33 ~$^{+0.27}_{-0.28}$ \%,  the mean  number of photons impacting PMT 
is obtained. This is the first step for the  PDE calculation.

\paragraph{G-APD fit procedure}\label{sec:G-APDfit}

\begin{figure}[t]
\begin{center}
\includegraphics[width=0.47\linewidth,clip]{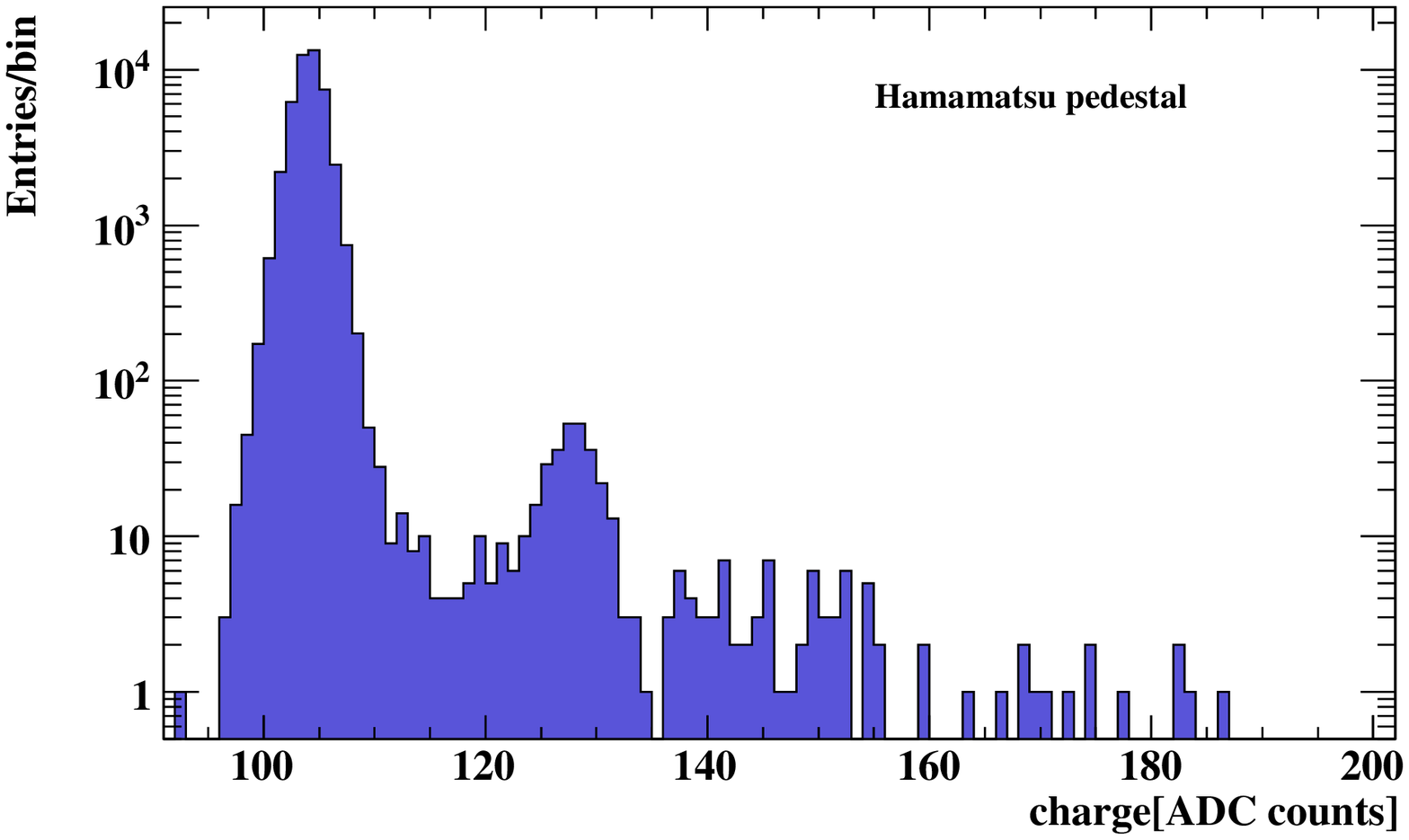}
\includegraphics[width=0.47\linewidth,clip]{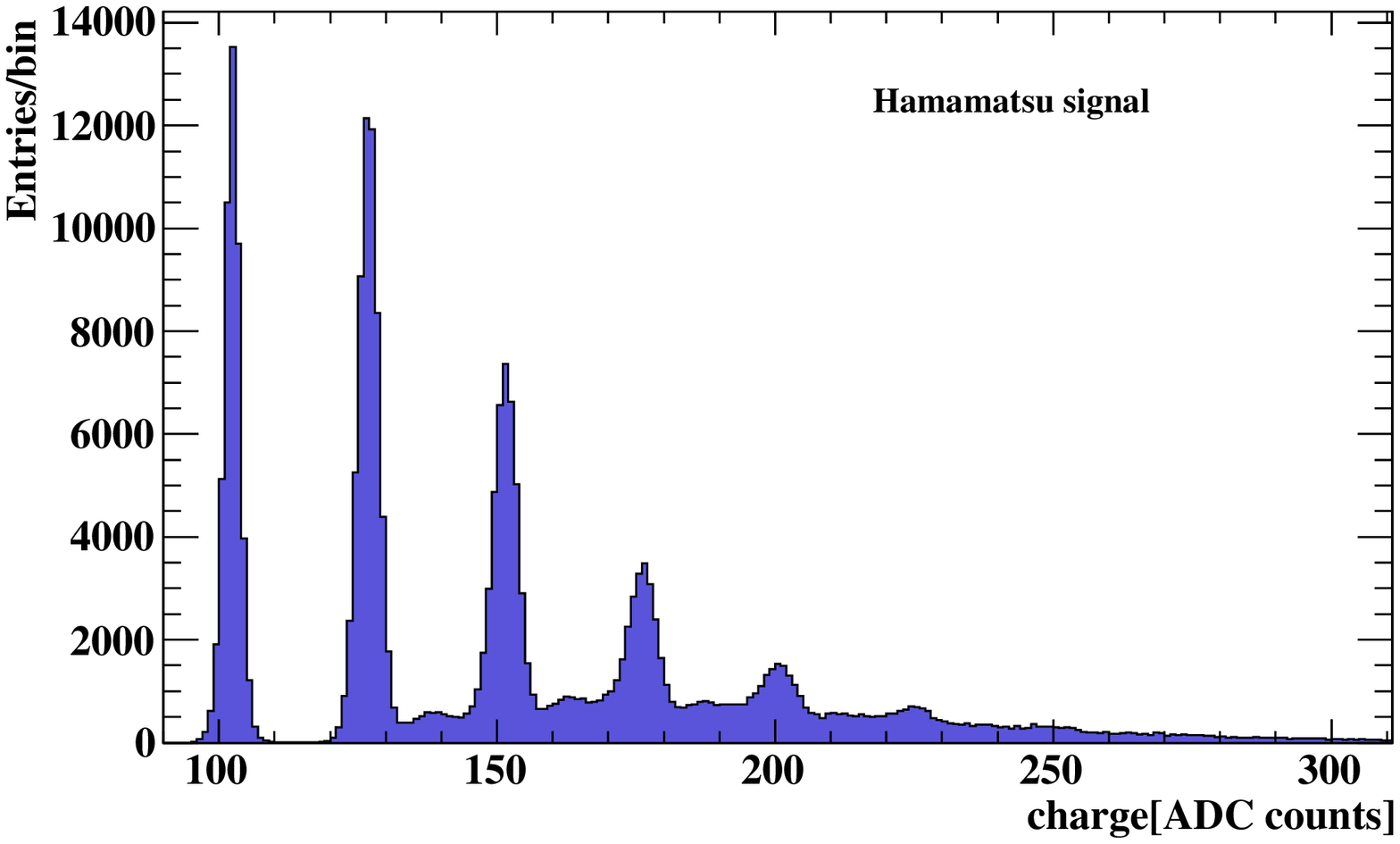}
\end{center}
\caption{ HAMAMATSU G-APD: noise  (left) and  signal (right) spectra.
The signal is  obtained  by short, low-intensity LED  flashes  at $\lambda_{\rm w}$ = 450 $\pm$ 3 nm.
 Each peak corresponds to a number of detected photons. Binwidth=1.}
\label{fig:HAM1}
\end{figure}

Figure ~\ref{fig:HAM1} shows  typical  G-APD HAMAMATSU response spectra, on the left  for no light (noise) and 
  on the right for low-intensity light (signal).

\par The noise spectrum shows beyond  the first peak (pedestal)  other peaks which are due to
 thermogeneration.
 Since their contribution to the total noise is about 2\%  of the pedestal,  the thermogeneration effect  can be neglected  in the signal fit.
Moreover since
 the probability of two or more thermogeneration counts within the  gate is negligible, the cross-talk effect can be estimated
from the  ratio of the numbers of events in the $\rm 3^{rd}$ 
 and  $\rm 2^{nd}$ peak.
 This value ($\sim 20\%$) depends on the gate length
(in our case 65 ns) and on the operation voltage, $\rm{V_{bias}}$ (Sec. \ref{sec:principle}).

\par In the signal spectrum the first peak (pedestal) corresponds to the noise,
 the second one is the G-APD response when
 exactly one photon is detected (one pixel is fired).
It is impossible to identify which one of  the pixels was fired, since they are all connected to the same output and they have 
similar responses. 
If n  pixels are fired, the sum of n charges is recorded  at the position of the $\rm(n+1)^{\rm th}$ peak, $ \mu_n=\mu_0+{\rm n\times g }$.

These peaks, as shown in Fig.~\ref{fig:HAM1} (right), are equidistant, at a distance  
 determined by the gain factor, g.
The  enhancements which show up  
 between the peaks, were attributed to  after-pulses (AP) and were considered in the fit.

\begin{figure}[t]
\begin{center}
\includegraphics[width=0.45\linewidth,clip]{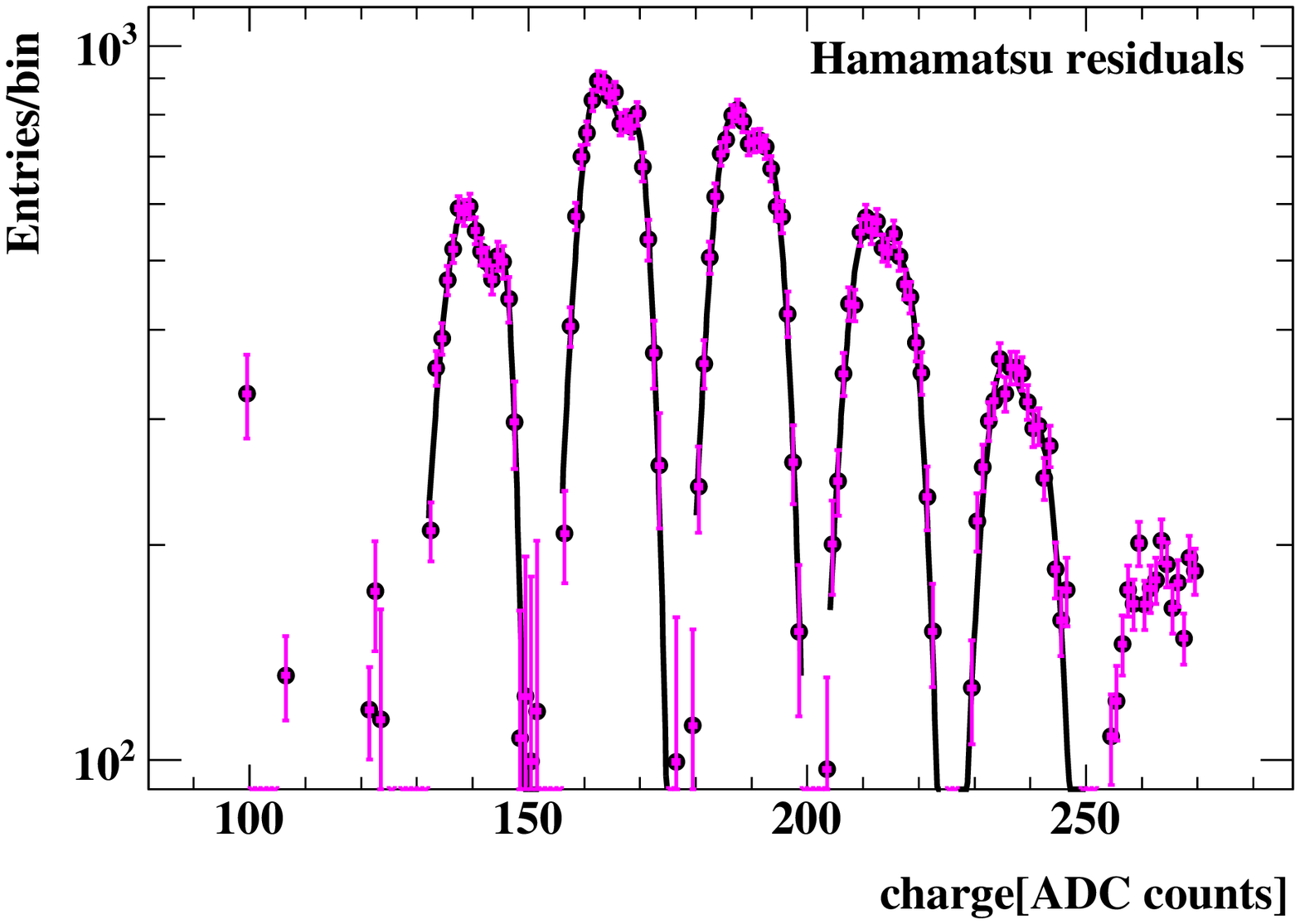}
\includegraphics[width=0.52\linewidth,clip]{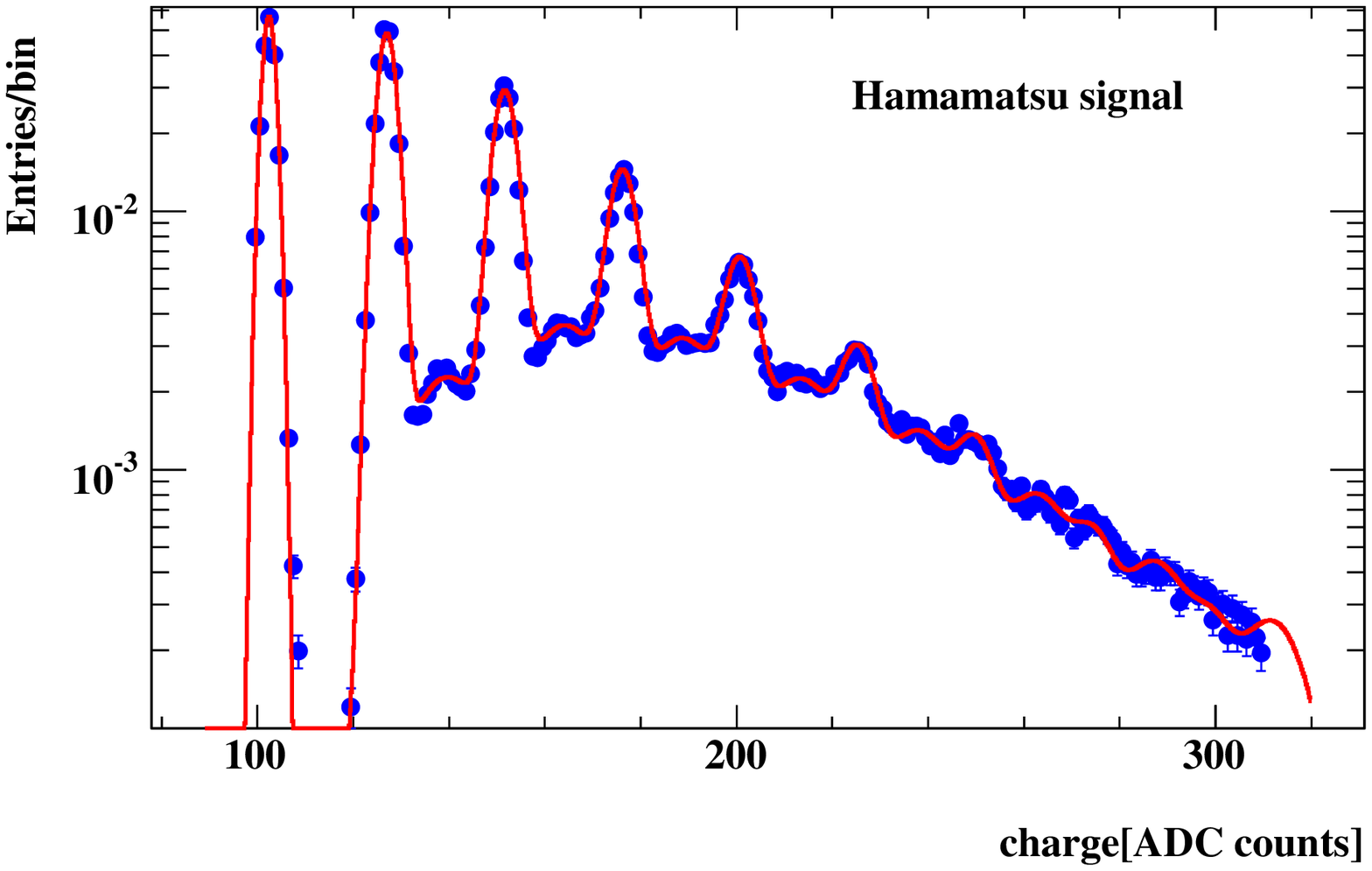}
\end{center}
\caption{Residual distribution after  subtraction of best-fitted  
gaussians from the signal spectrum (left). The signal  spectrum fitted 
with the after-pulse corrections (right). From this fit, the probability,
$\rm P_{\rm n}^{\rm 0}$,  to get n cells fired  and
the  probability, $\rm P_{\rm{AP}}$, to get an after-pulse from one cell are derived (in this example $\rm P_{\rm{AP}}= 0.16 \pm 0.01 $).
Binwidth=1.}
\label{fig:AP}
\end{figure}

 To identify  the  most appropriate 
function describing this  effect  the main peaks, each fitted to a  gaussian distribution, were subtracted 
from the spectrum of Fig.~\ref{fig:HAM1} (right). As a result,  a sequence of peaks, Fig.~\ref{fig:AP} (left), was obtained, each  one well described
as a sum of two gaussians (for this G-APD and for the chosen gate length). This suggested  to write a new fitting function P(x) as follows.

If  $\rm P_n^0$ is the probability to get initially n cells fired (the poissonian n term in  Eq.~\ref {eq:eq3}),
the probability to record an x charge, P(x)=N(x)/N 
can be written as sum of three terms
representing respectively the probability of  an x charge when  n  cells are fired 1) without AP, $\rm P_n^{\rm{noAP}}$;
2) when only  one AP occurred, $\rm P_n^{\rm{AP1}}$; 3) when two APs occurred, $\rm P_n^{\rm{AP2}}$:
\begin{equation}
\rm P(x) = \sum_{n} P_n^{\rm{noAP}}\times Gauss(\mu_n, \sigma_n) + 
\rm P_n^{\rm{AP1}}\times Gauss(\mu_n+\delta_1, \sigma_1) +
\rm P_n^{\rm{AP2}}\times Gauss(\mu_n+\delta_2, \sigma_2), 
\label{eq:tot_prob}
\end{equation}

\noindent

\noindent where $\delta_{1,2}$ are  the distances of the first and second  after-pulse gaussian 
 from the  nearest main peak  and  $\sigma_{1,2}$ the widths.
 The  probabilities  ${\rm P_n^{noAP}}$,  ${\rm P_n^{AP1}}$ and ${\rm P_n^{AP2}}$  
 are all functions  of a single parameter ${\rm P_{AP}}$, the probability to get an AP from one cell, and can be written
as:

\begin{align}
\nonumber
\rm  P_n ^{noAP}    &= \rm P_n^0\times(1-P_{AP})^n\;, \\ \nonumber 
\rm  P_n^{\rm{AP1}} &= \rm P_n^0\times(1-(1-P_{\rm{AP}})^n)\times (1-P_{\rm{AP}})^n, \\ \nonumber
\rm  P_n^{\rm{AP2}} &= \rm P_n^0\times(1-(1-P_{\rm{AP}})^n)\times(1-(1-P_{\rm{AP}})^n).  \nonumber
\end{align}


\noindent If the  probability of after-pulses is zero, the  Eq.~\ref{eq:tot_prob} is equivalent to Eq.~\ref {eq:eq3}.

\begin{figure}[t]
\begin{center}
\includegraphics[width=0.47\linewidth,clip]{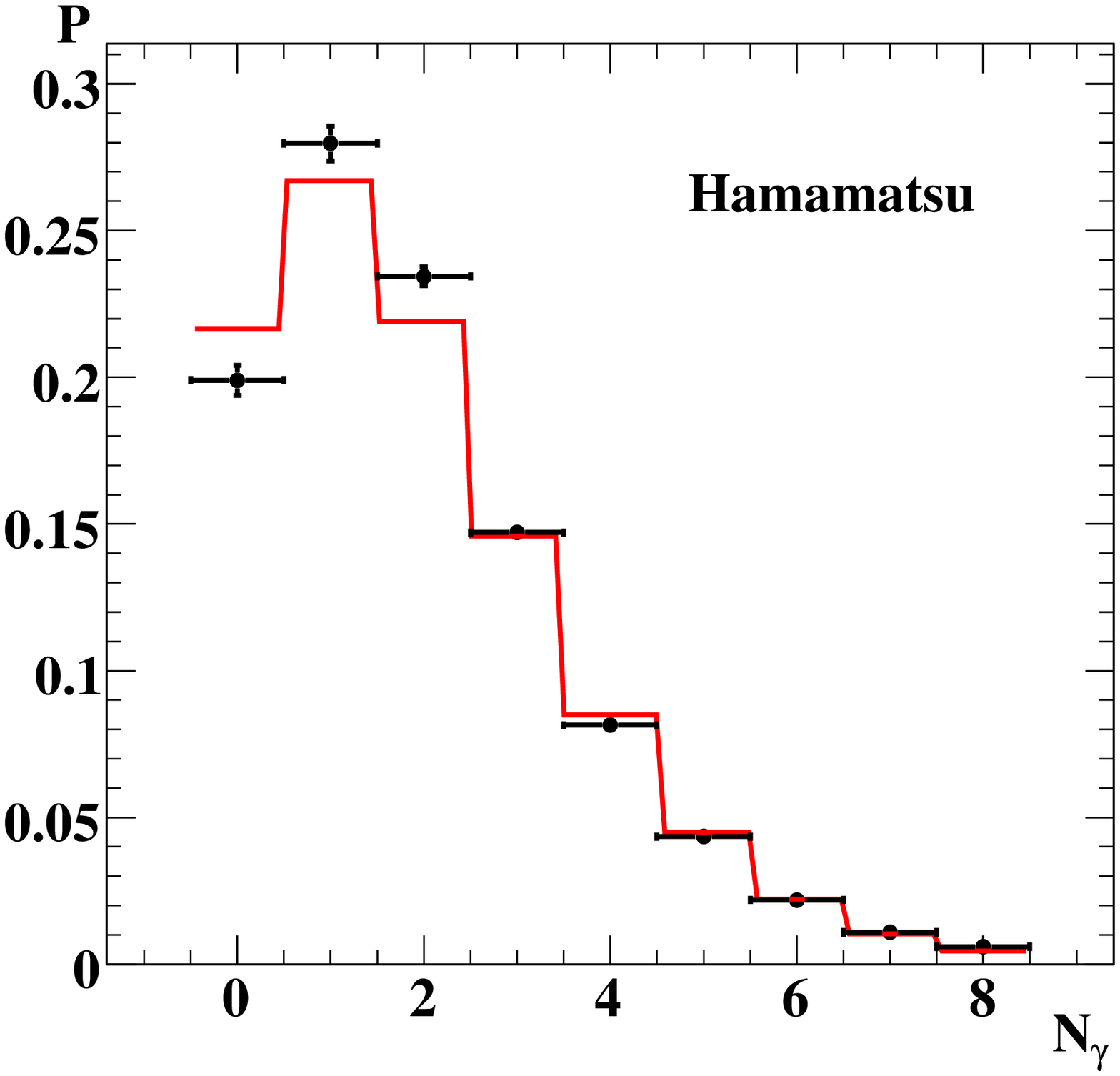}
\includegraphics[width=0.47\linewidth,clip]{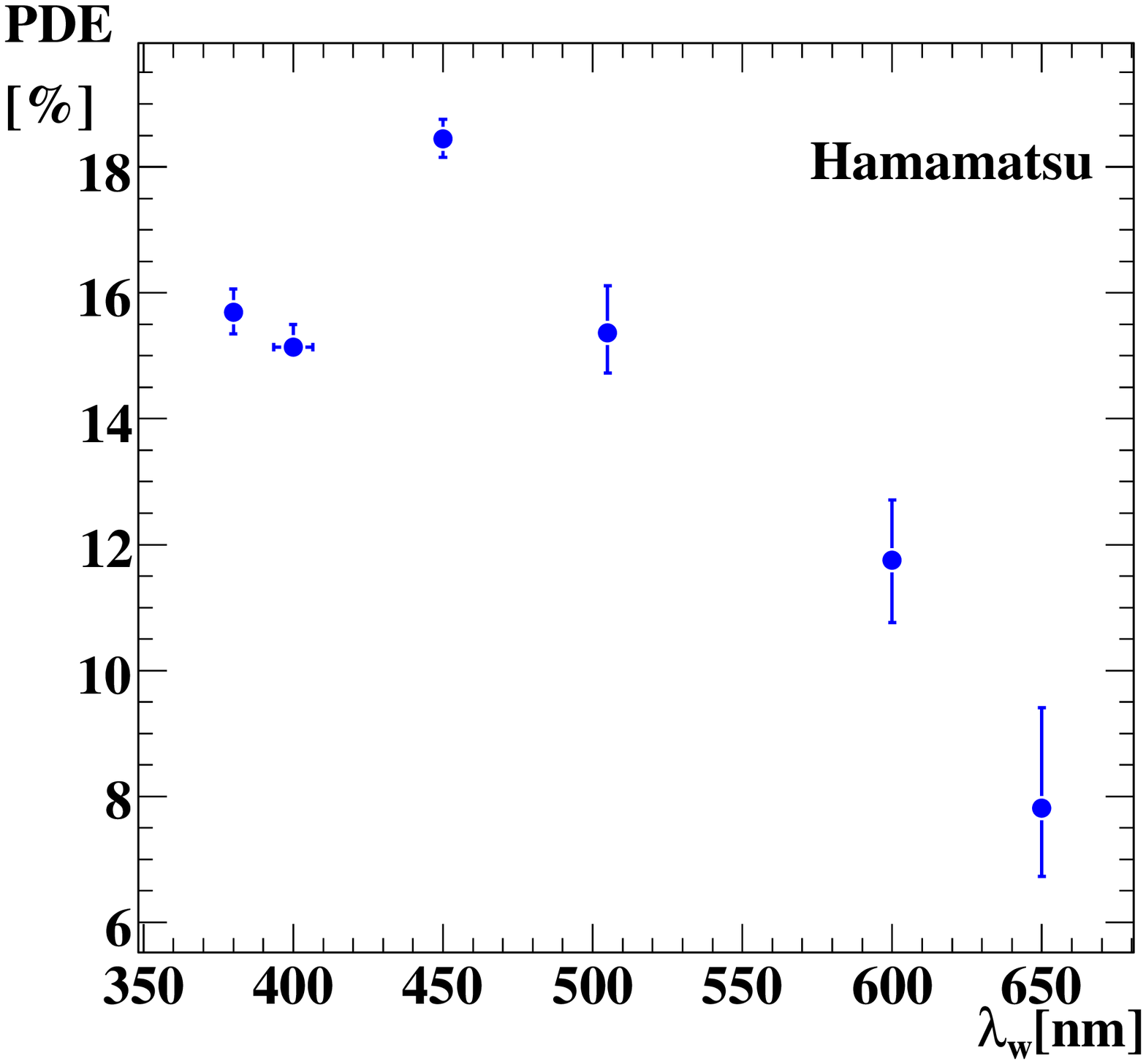}
\end{center}
\caption{Right: distribution of probability, P, to observe $\rm N_{\gamma}$ photons. 
The data, black dots, are the
$\rm P_n^{0}$  obtained by   fitting   the signal spectrum  (Fig.~\ref{fig:AP}, right).
 The red line represents the fit of  these data with Eq. \ref{eq:Poisson}.
 In this example, the fitted values are:
 the mean  number of G-APD detected photons, $\rm N_{\gamma}^{G-APD}=1.55 \pm 0.02 $,
 and the cross-talk probability, $\varepsilon=0.20\pm 0.01$. Binwidth=1.
Left: The photon detection efficiency, PDE, for the HAMAMATSU G-APD as a function of the light wavelength, $\lambda_w$.}

\label{fig:XT}
\end{figure}

 Equation ~\ref{eq:tot_prob} was fitted to the spectrum of Fig.~\ref{fig:HAM1} (right)
and the fit result is  shown in Fig.~\ref{fig:AP} (right) as a continuous line. Among the  free parameters, we obtain the $\rm P_n^{0}$ values 
 as displayed in Fig.~\ref{fig:XT} (black dots) and  the after-pulse probability, ${\rm P_{\rm AP}}\ =\  16\%$. 


 The  $\rm P_n^{0}$ values  should be distributed according to the Poisson statistics
with a mean number of photons detected,  $\rm N_{\gamma}^{G-APD}$ (corresponding to the Poisson's distribution parameter $\lambda$).
 However this distribution is distorted by the presence of cross-talk.
If the cross-talk probability is  $\varepsilon \neq 0$, then the n=1 bin
 ($2^{\rm nd}$ peak in Fig.~\ref{fig:AP} (right)), 
contains  only  events with   one initially fired cell without any cross-talk, with a
probability P(1)= $\rm P_1^{0}(1-\varepsilon)$. 
The n=2 bin  ($3^{\rm rd}$  peak) is filled either when two cells are  
 initially fired  without cross-talk or when  one  cell is fired  accompanied by a second cell fired by 
cross-talk: $\rm P(2)=P_2^{0}(1-\varepsilon)^2+P_1^{0}\varepsilon(1-\varepsilon)$.
 Finally the probability
to observe n  fired cells  P(n), can be written:
\begin{equation}
\rm
P(n)= \sum_{j=1}^n P_j^{0}(1-\varepsilon)^j\varepsilon^{n-j}\binom{n-1}{j-1}.
\label{eq:Poisson}
\end{equation} 
 The data in  Fig.~\ref{fig:XT} (left)  were fitted using Eq.~\ref{eq:Poisson};
 the red line represents  the result of the fit. For these data, the  value obtained for the cross-talk 
probability was $\varepsilon=0.20\pm 0.01$; the value of the  mean number of detected  
photons  is   $\rm N_{\gamma}^{G-APD}=1.55 \pm 0.02$.
In general, for any pair of measurements on G-APD sample, and on the reference PMT, we have:
\begin{equation}
\rm  PDE= \frac{N_{\gamma}^{G-APD}}{N_{\gamma}^{PMT}\cdot \epsilon_{eval}^{ ref}} 
\label{eq:PDE_formula}
\end{equation}

\par With  the mean number of  photons detected by G-APD,
$\rm N_{\gamma}^{G-APD}$ and  by PMT, the reference detector, $\rm N_{\gamma}^{PMT}$,
it is possible to evaluate the photon detection efficiency. In our example, 
 the  PDE value is  $18.36 \pm  0.26 \ \%$ (statistical error only).

 The final value of PDE is obtained as mean of various measurements at different light intensities,
with different  fiber configurations, taking into account the FWHM of the light source and the filter bandwidth. 

The procedure and results described in this section for G-APD HAMAMATSU were as well
applied to the others samples, with minor modifications.
In the following sections these results are discussed.

\section{Photon detection efficiency: Hamamatsu sample} \label{sec:PDE_HAMA}

Based on the  fitting procedure of Sec. \ref{sec:Fit}, it is possible to calculate
for each measurement the corresponding  PDE value. The measurements have been made at different light
intensities and with different optical layouts to  evaluate  possible sources of systematics.
The systematic error  due to optical contact, excluding a statistical contribution, is $\sim$ 1\% in absolute value.

 At this point, it is possible to evaluate the  photon detection efficiency of the device  as the mean of all available measurements.
 The results are shown in Fig. \ref{fig:XT} (right). The horizontal  error  bar  is originated from the light source FWHM or filter bandwith; 
its impact on PDE evaluation
  gives the vertical error bar.
\section{Photon detection efficiency: CPTA sample} \label{sec:PDE_CPTA}

\begin{figure}[t]
\begin{center}
\includegraphics[width=0.47\linewidth,clip]{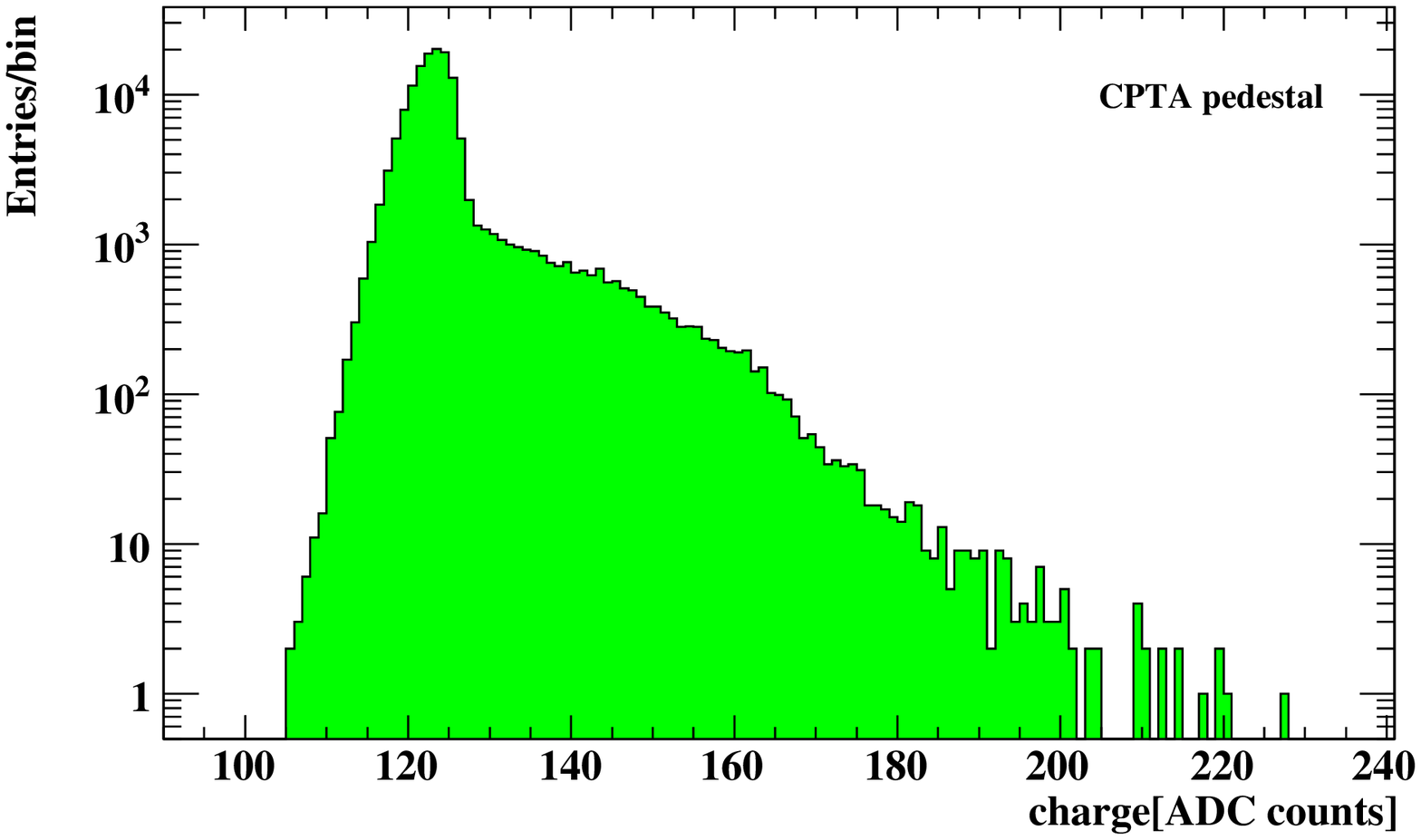}
\includegraphics[width=0.47\linewidth,clip]{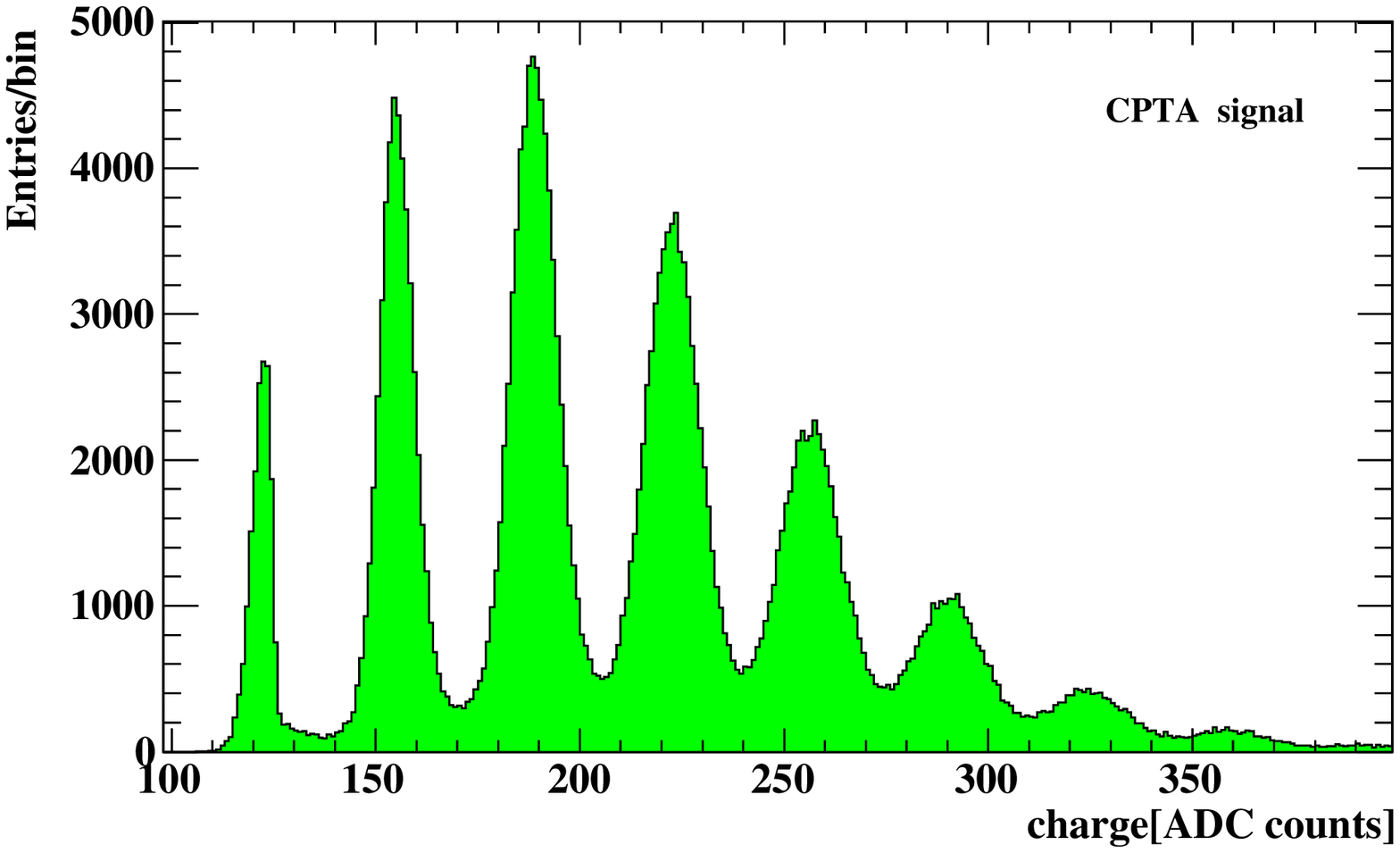}

\end{center}
\caption{The noise  (left) and signal (right) spectra obtained with the
  CPTA G-APD illuminated by short, low intensity LED  flashes  at $\lambda_w$ = 600 $\pm$ 3 nm. Each peak corresponds to a  number of detected photons. Binwidth=1.}
\label{fig:CPTA1}
\end{figure}

The  noise and the signal spectra obtained with the CPTA sample, as shown in Fig.~\ref{fig:CPTA1}   for $\lambda_w$ = 600 $\pm$ 3 nm, reflect the difference on the construction characteristics.
With respect to the HAMAMATSU device  Fig.~\ref{fig:CPTA1} (left) shows that with the chosen gate 
the thermogeneration is at a level of $\sim$ 30\% , and Fig.~\ref{fig:CPTA1} (right) shows that
 the after-pulses 
have lower impact  on the  single photon spectra.
The  cross-talk effect is much lower.
The fit procedure adopted for the CPTA device is slightly different from that described in  Sec.~\ref{sec:Fit}. 
The thermogeneration contribution was not included in the fit, the AP terms ( Eq.~\ref{eq:tot_prob}) were not directly related to an  after-pulse effect but accounted  also for the termogeneration and  other effects.

\begin{figure}[t]
\begin{center}
\includegraphics[width=0.47\linewidth,clip]{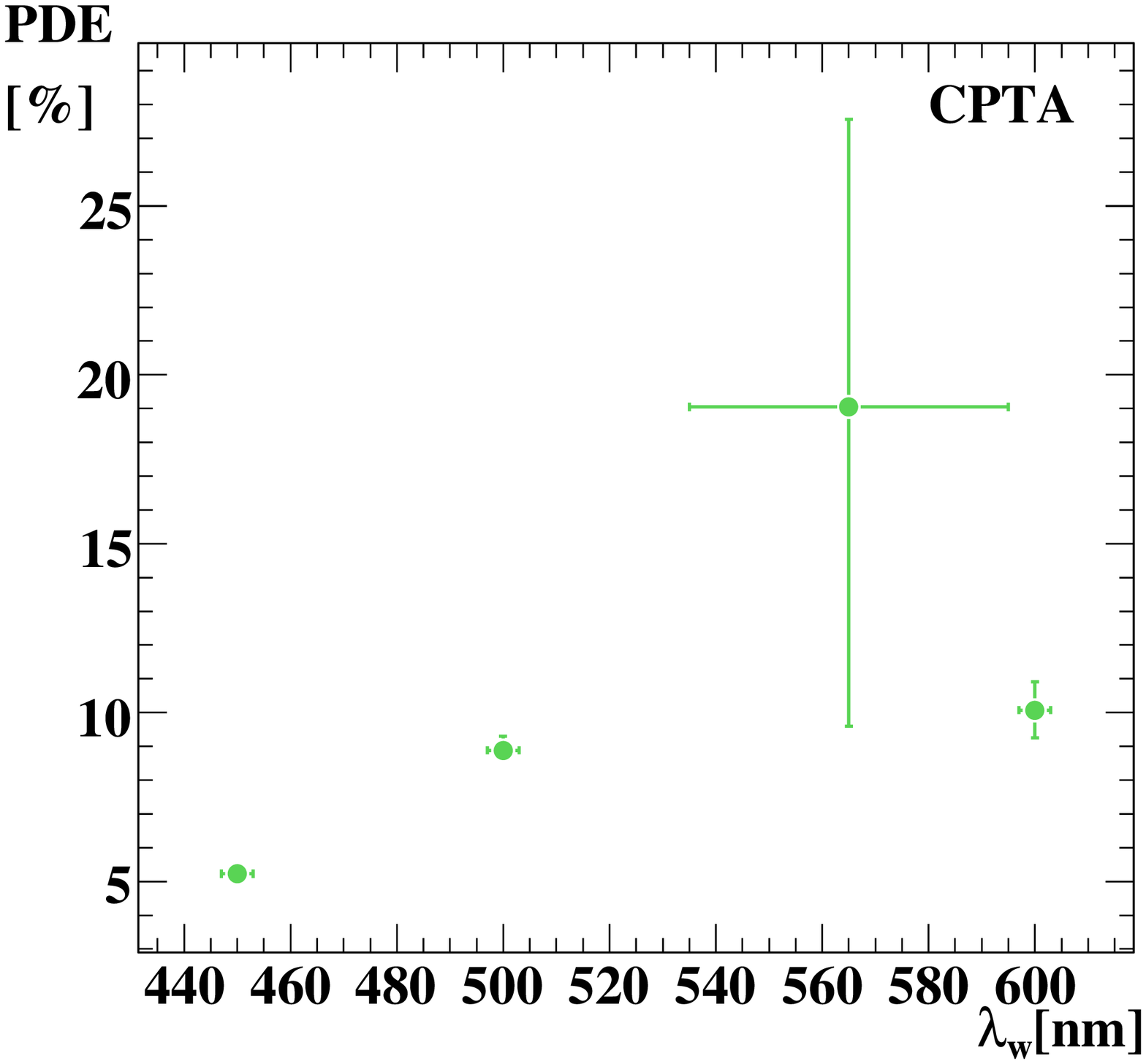}
\includegraphics[width=0.47\linewidth,clip]{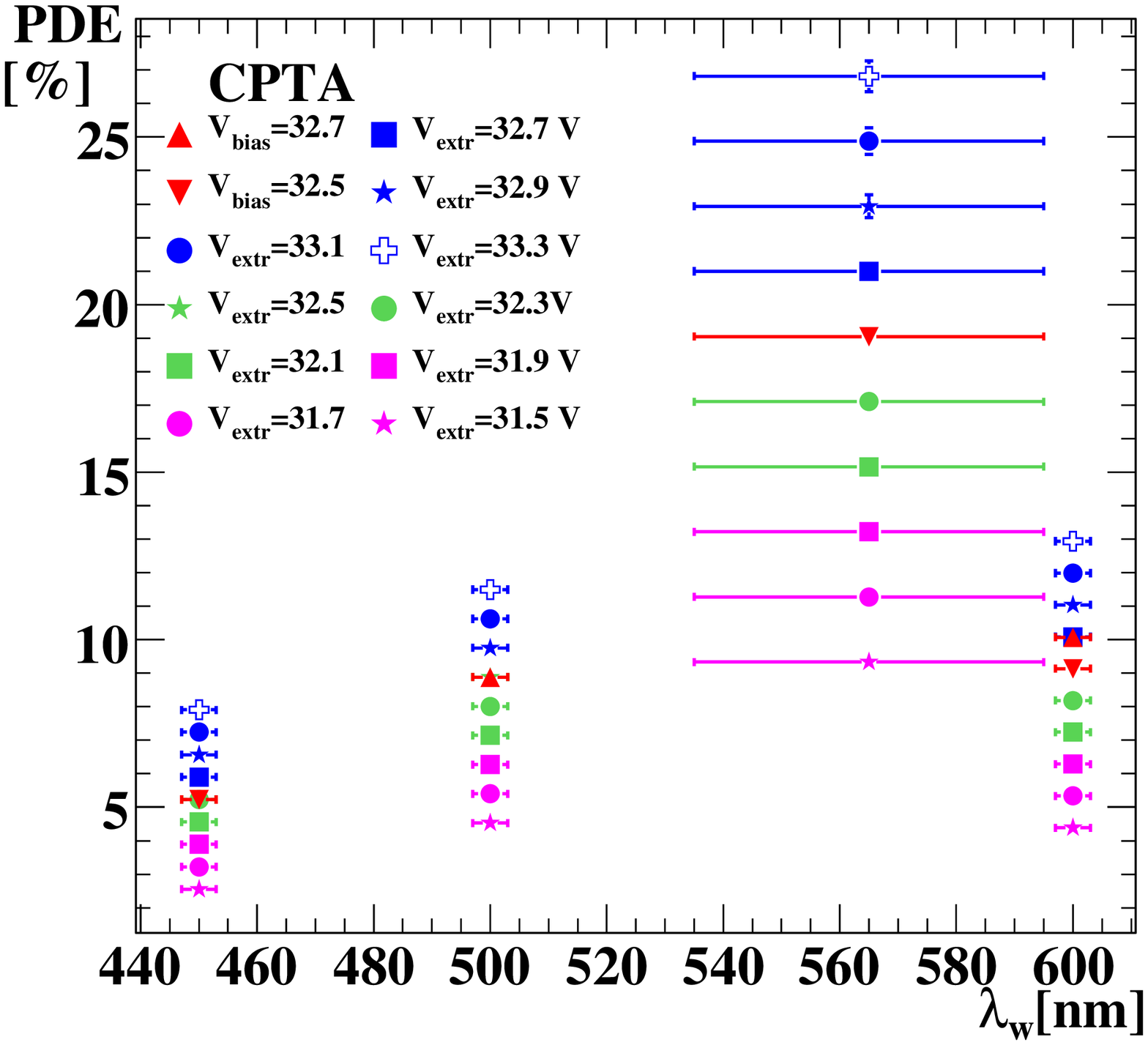}

\end{center}
\caption{
Left: the photon detection efficiency, PDE, for CPTA G-APD as a function of the light wavelength, $\lambda_w$.
The measurements at 450 nm, 565 nm, 600 nm have been recorded
at $\rm{V_{bias}}$=32.5 V, those at 500 nm, 600 nm at $\rm{V_{bias}}$ = 32.7 V. The $\lambda_w$ = 600 nm measurements at 
the two bias voltages  are indistinguishable. Right: the photon detection efficiency, PDE, for the CPTA detector as a function of the wavelength,  $\lambda_w$,
as measured (red points) and extrapolated (green, blue, magenta points).}
\label{fig:PDE_CPTA}
\end{figure}

The PDE results for other wavelengths are shown in Fig. \ref{fig:PDE_CPTA} (left). 
The horizontal error  bar is originated from the light source FWHM or  filter bandwidth, the vertical one from
 its impact on the PDE evaluation. The measurements have been recorded at  $\rm{V_{bias}}$=32.5 V for $\lambda_w$=
 450 nm, 565 nm, 600 nm
and  $\rm{V_{bias}}$=32.7 V for $\lambda_w$= 500 nm and 600 nm.
The measurement at $\lambda_w$= 600 nm has been recorded at both voltages.
\subsection{Parameter dependence on overvoltage: PDE}

The  photon detection efficiency, $\rm{PDE_{rel}}$, relative  to the one at the reference voltage as a function of $\rm{V_{bias}}$ has been 
recorded at different wavelengths. From a straight line fits ($\rm{PDE_{rel}} =p_0+p_1 \rm{V_{bias}}$) 
these distributions the  values for the angular coefficients, $\rm p_1$$[\rm V^{-1}]$ are derived and
 the   PDE value   as a function of $\lambda_w$ at various
values of ${\rm V_{bias}}$ is obtained, Fig.~\ref{fig:PDE_CPTA} (right) . 
 The validity of the  method is tested  with  the measurements at 600 nm, taken
at two different bias voltages.

\subsection{An improved fit procedure } \label{sec:FURTHER_CPTA}

The CPTA G-APD measurements described above have two
peculiar aspects that deserve more attention: 
the thermogeneration effect is not negligible (Fig.~\ref{fig:CPTA1} (left) and the after-pulse effect requests a dedicated treatment. We refere to ~\cite{Gentile:2010nuovo}, for a detailed description of the method.

As a first step  the termogeneration spectrum  shape is determined as a deviation  of the noise spectrum, Fig.~\ref {fig:CPTA1} (left), from a gaussian. A combination of two exponential functions results into an appropriate description of this effect.

As second step the fit of the signal  pedestal 
is not performed with a simple gaussian as above, but includes also the fit function determined in the first step and describing the termogeneration. 

\begin{wrapfigure}{l}{0.5\columnwidth}
\centerline{\includegraphics[width=0.45\columnwidth]{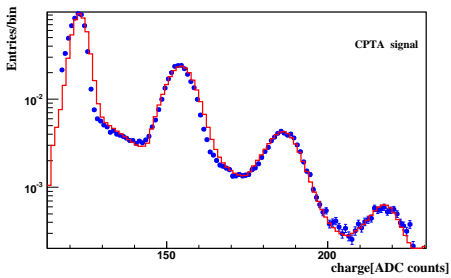}}
\caption{The CPTA G-APD signal spectrum (at $\lambda_{\rm w}$=450 $\pm$ 3 nm) fitted 
with the improved fit procedure, Sec.~\ref{sec:FURTHER_CPTA}.
}\label{fig:CPTA_asym}
\end{wrapfigure}

 Further improvement  in the signal fit  is  achieved,  if the after-pulse effect is modeled  for a single fired pixel, as suggested in Sec. \ref{sec:principle}. The results of this improved method are shown in Fig.~\ref {fig:CPTA_asym}. 


\noindent Using this  fit procedure it is possible to derive a value
 of the cross-talk probability  and the average obtained on measurements
performed at many light intensities is   $\leq 1.7\%$. 

 
Similar measurements were  recorded with a IRST device.  
Again, the noise and signal spectra reflect the difference on construction and design.
 From these spectra
we have concluded that the specimen used wasn't suitable for PDE measurements.

\section{Conclusion} \label{sec:Conclusion}

Measurements of G-APD response to low intensity light were recorded to
determine the main characteristics of three different samples,
HAMAMATSU (S10362-11-025U), CPTA, IRST. An accurate  method to fit  the response
has been realized, to evaluate  photon detection efficiency,
cross-talk, gain, and after-pulse probability. A calibrated reference detector has been used
for the photon detection efficiency measurements in a wavelength range 
 between 380 and 650 nm, see Ref.~\cite{Gentile:2010nuovo}.
\par The measurements were carried on in the same experimental condition (setup, thermal insulation...)
to compare the properties of various samples.
This information is an important issue in the near future
to identify the most suitable device for a particular
application in high-energy physics calorimeters, astroparticle physics and  medical environment.

The PDE measurements  provided are consistent with those quoted from the manufacturers with a {\it caveat} 
due to the different method of measurement (the current method, very often used by manufacturers, includes after-pulse and cross-talk effects).

\section{Acknowledgments}\label{sec:Akno}

 We would like to thank Profs. M. Danilov and  E. Tarkovsky for the  samples provided  and clarifying discussions,
 Dr. I. Polak  for having provided a fundamental part of electronics,
and  Prof. Maria Fidecaro for comments to the manuscript.
\section{Bibliography}

\begin{footnotesize}


\end{footnotesize}


\end{document}